\documentclass[prl,twocolumn,showpacs,superscriptaddress,preprintnumbers,amsmath,amssym]{revtex4-1}

\usepackage[latin1]{inputenc}
\usepackage{bm}
\usepackage{multirow,amssymb,amsbsy,amsmath}
\usepackage{graphicx}
\usepackage{verbatim}
\makeatletter
\usepackage{pifont}

\usepackage{epstopdf}
\usepackage[justification=RaggedRight,singlelinecheck=false]{caption}
\usepackage[justification=centering]{subfig}

\makeatother

\newcommand{\mS}{\mathcal{S}(\mathcal{H})}

\newcommand{\mH}{\mathcal{H}}
\newcommand{\mE}{\mathcal{E}}
\newcommand{\mD}{\mathcal{D}}
\newcommand{\mN}{\mathcal{N}}

\begin{document}

\title{Locality and universality of quantum memory effects}

\author{B.-H. Liu}
\thanks{These authors contributed equally to this work.}
\affiliation{Key Laboratory of Quantum Information, University of Science and 
Technology of China, CAS, Hefei, 230026, China}

\author{S. Wi{\ss}mann}
\thanks{These authors contributed equally to this work.}
\affiliation{Physikalisches Institut, Universit\"at Freiburg,
Hermann-Herder-Stra{\ss}e 3, D-79104 Freiburg, Germany}

\author{X.-M. Hu}
\affiliation{Key Laboratory of Quantum Information, University of Science and 
Technology of China, CAS, Hefei, 230026, China}

\author{C. Zhang}
\affiliation{Key Laboratory of Quantum Information, University of Science and 
Technology of China, CAS, Hefei, 230026, China}

\author{Y.-F. Huang}
\affiliation{Key Laboratory of Quantum Information, University of Science and 
Technology of China, CAS, Hefei, 230026, China}

\author{C.-F. Li}
\email{cfli@ustc.edu.cn}
\affiliation{Key Laboratory of Quantum Information, University of Science and 
Technology of China, CAS, Hefei, 230026, China}

\author{G.-C. Guo}
\affiliation{Key Laboratory of Quantum Information, University of Science and 
Technology of China, CAS, Hefei, 230026, China}

\author{A. Karlsson}
\affiliation{Turku Centre for Quantum Physics, Department of Physics and 
Astronomy, University of Turku, FI-20014 Turun yliopisto, Finland}

\author{J. Piilo}
\affiliation{Turku Centre for Quantum Physics, Department of Physics and 
Astronomy, University of Turku, FI-20014 Turun yliopisto, Finland}

\author{H.-P. Breuer}
\email{breuer@physik.uni-freiburg.de}
\affiliation{Physikalisches Institut, Universit\"at Freiburg,
Hermann-Herder-Stra{\ss}e 3, D-79104 Freiburg, Germany}

\date{\today}

\begin{abstract}
Recently, a series of different measures quantifying memory effects in
the quantum dynamics of open systems has been proposed. Here, we derive
a mathematical representation for the non-Markovianity measure based on the
exchange of information between the open system and its environment which
substantially simplifies its numerical and experimental determination, and fully
reveals the locality and universality of non-Markovianity in the quantum
state space. We further illustrate the application of this representation by means of 
an all-optical experiment which allows the measurement of the degree of memory
effects in a photonic quantum process with high accuracy.
\end{abstract}


\maketitle

In recent years the problem of characterizing non-Markovian dynamics
in the quantum regime has initiated an intense debate. A series of diverse
definitions along with measures of quantum memory effects have been proposed,
invoking many different mathematical and physical concepts and techniques.
Examples are characterizations of non-Markovianity in terms of deviations from
a Lindblad semigroup \cite{Wolf}, of the divisibility
of the dynamical map \cite{RHP}, of the dynamics of entanglement \cite{RHP} and 
correlations \cite{Luo} with an ancilla system, and of the Fisher information
\cite{Sun}.

In this work we focus on the measure of non-Markovianity introduced in 
Refs.~\cite{BLP,BLP2} which defines non-Markovianity through the backflow of 
information from the environment to the open system. This information 
backflow is characterized by an increase of the distinguishability of 
time-evolved quantum states. The distinguishability of two quantum states 
$\rho_1$ and $\rho_2$ is quantified by their trace distance 
$\mD(\rho_1,\rho_2) = \frac{1}{2} {\mathrm{Tr}} \left| \rho_1-\rho_2 \right|$ 
\cite{Fuchs,Hayashi,Nielsen}.
We assume that the open system Hilbert spaces $\mH$ is finite dimensional. The 
corresponding space of physical states, 
represented by the convex set of positive operators with unit trace, will be denoted 
by $\mS$. We further assume that the time evolution of the open system can be 
described by a 1-parameter family $\Phi$ \cite{Breuer2007} of completely positive 
and trace preserving dynamical maps $\Phi_t$, i.e. 
$\Phi = \{ \Phi_t \mid 0\leq t \leq T, \; \Phi_0 = I\}$. The non-Markovianity
measure can then be defined as
\begin{equation}
 \mathcal{N}(\Phi) = 
 \max_{\rho_1\perp\rho_2} \int_{\sigma>0} dt ~ 
 \sigma(t,\rho_1,\rho_2), \label{eq:measure-ortho}
\end{equation}
where
\begin{equation}
 \sigma(t,\rho_1,\rho_2) \equiv \frac{d}{dt} 
 \mathcal{D}(\Phi_t(\rho_1),\Phi_t(\rho_2))
\end{equation}
denotes the time derivative of the trace distance between the pair of states
at time $t$. In Eq.~\eqref{eq:measure-ortho} the time integral is extended
over all time intervals in which this derivative is positive, and the maximum
is taken over all pairs of orthogonal initial states $\rho_1\perp\rho_2$.
This measure for non-Markovianity was originally defined in \cite{BLP} in 
terms of a maximization over all pairs of quantum states in $\mS$. However, as
demonstrated in Ref. \cite{OptimalStates} the maximization can be restricted to
pairs of orthogonal initial states.
We recall that two quantum states $\rho_1$ and $\rho_2$ are said to be 
orthogonal if their supports, i.e. the subspaces spanned by 
their nonzero eigenvalues are orthogonal which is equivalent to 
$\mathcal{D}(\rho_1,\rho_2)=1$ \cite{Nielsen}. This implies that optimal state pairs 
exhibiting a maximal backflow of information during their time evolution are initially 
distinguishable with certainty, and thus represent a maximal initial information 
content.

Although the orthogonality of optimal states greatly simplifies the
mathematical representation of the non-Markovianity measure, its determination 
still requires the maximization over pairs of quantum states. Here, we derive a
much simpler representation for the measure which is particularly relevant for its 
experimental realization since it only requires a local maximization over single 
quantum states, the second state being an arbitrary fixed reference state taken 
from the interior of the state space. This representation will further be employed in 
an all-optical experimental setup for the measurement of the non-Markovianity of a 
photonic process.

To formulate our main theoretical result we first define 
$\mathring{\mathcal{S}}(\mH)$
to be the interior of the state space, i.e. the set of all quantum states $\rho_0$
for which there is an $\varepsilon>0$ such that all Hermitian operators $\rho$ with
unit trace satisfying $\mathcal{D}(\rho,\rho_0)\leq\varepsilon$ belong to
$\mS$. We further define
\begin{equation}
 \mE_0(\mH) =\{ A \mid A\neq0, \, A=A^\dag, \, 
 \mathrm{Tr}\, A=0 \}
\end{equation}
to be the set of all nonzero, Hermitian and traceless operators on $\mH$.
Considering any fixed reference state $\rho_0\in\mathring{\mathcal{S}}(\mH)$ we 
can now introduce a particular class of subsets of the state space:
A set $\partial U(\rho_0)\subset\mS$ not containing $\rho_0$ is called an 
enclosing surface of $\rho_0$ if and only if for any operator $A\in\mE_0(\mH)$ 
there exists a real number $\lambda>0$ such that 
\begin{equation} \label{def-U}
 \rho_0+\lambda A\in \partial U(\rho_0).
\end{equation}
Note that by definition $\rho_0$ itself is not contained in $\partial U(\rho_0)$
and that the full set $\partial U(\rho_0)$ is part of the state space. It can be easily 
seen that any state from the interior of the state space has an enclosing surface. 
For example, since $\rho_0$ is an interior point of the state space there
is an $\varepsilon>0$ such that the set of states $\rho$
defined by $\mathcal{D}(\rho,\rho_0)=\varepsilon$ represents a spherical 
enclosing surface with center $\rho_0$. However, an enclosing surface 
$\partial U(\rho_0)$ can have an arbitrary geometrical shape, the only requirement 
being that it encloses the reference  state in all directions of state space. An 
example is shown in Fig.~1(a). Using these definitions, we can now state our 
central result.

\paragraph{Theorem.}
Let $\rho_0\in\mathring{\mathcal{S}}(\mH)$ be any fixed state of the interior of 
the state space and $\partial U(\rho_0)$ an arbitrary enclosing surface of 
$\rho_0$. For any dynamical process $\Phi$, the measure for quantum 
non-Markovianity defined by Eq.~\eqref{eq:measure-ortho} is then given by 
\begin{equation}
  \mathcal{N}(\Phi) = 
  \max_{\rho\in\partial U(\rho_0)} \int_{\bar\sigma>0} dt ~ \bar\sigma(t,
  \rho,\rho_0), \label{eq:new measure}
\end{equation}
where
\begin{equation}
 \bar\sigma(t,\rho,\rho_0) \equiv \frac{\frac{d}{dt}\mathcal{D}(\Phi_t(\rho),
 \Phi_t(\rho_0))}{\mathcal{D}(\rho,\rho_0)}
\end{equation}
is the derivative of the trace distance at time $t$ divided by the initial trace 
distance.

\paragraph{Proof.}
Let $\rho\in\partial U(\rho_0)$. Applying the Jordan-Hahn decomposition
\cite{Nielsen} to the operator $\rho-\rho_0$ one concludes that there exists an 
orthogonal pair of states $\rho_1$ and $\rho_2$ such that
\begin{equation} \label{relation}
 \rho_1-\rho_2=\frac{\rho-\rho_0}{\mD(\rho,\rho_0)},
\end{equation}
and, hence, we have
\begin{equation}
 \mD(\Phi_t(\rho_1),\Phi_t(\rho_2))
 = \frac{\mD(\Phi_t(\rho),\Phi_t(\rho_0))}{\mD(\rho,\rho_0)},
\end{equation}
by the linearity of the dynamical maps and the homogeneity of the trace distance. 
This shows that $\sigma(t,\rho_1,\rho_2)=\bar\sigma(t,\rho,\rho_0)$. It follows that 
the right-hand side of Eq.~\eqref{eq:new measure} is smaller than or equal to 
$\mN(\Phi)$ as defined by Eq.~\eqref{eq:measure-ortho}. Conversely, suppose 
$\rho_1,\rho_2$ are two orthogonal states. Since $\rho_1-\rho_2\in\mE_0(\mH)$, 
there exists $\lambda>0$ such that 
$\rho\equiv\rho_0+\lambda(\rho_1-\rho_2)\in\partial U(\rho_0)$, 
by definition of an enclosing surface. Thus, one obtains
$\rho_1-\rho_2=(\rho-\rho_0)/\lambda$. Since $\rho_1\perp\rho_2$ we find
$\mD(\rho,\rho_0)/\lambda=\mD(\rho_1,\rho_2)=1$ and, hence, 
$\lambda=\mD(\rho,\rho_0)$. Thus, we are again led to Eq.~\eqref{relation} and 
to $\sigma(t,\rho_1,\rho_2)=\bar\sigma(t,\rho,\rho_0)$. This shows that the 
measure $\mN(\Phi)$ as defined by Eq.~\eqref{eq:measure-ortho} is smaller than 
or equal to the right-hand side of Eq.~\eqref{eq:new measure} which thus 
concludes the proof.

\begin{figure}[tbh]
\centering
 \begin{minipage}[c]{9cm}
	\includegraphics[width=0.39\textwidth]{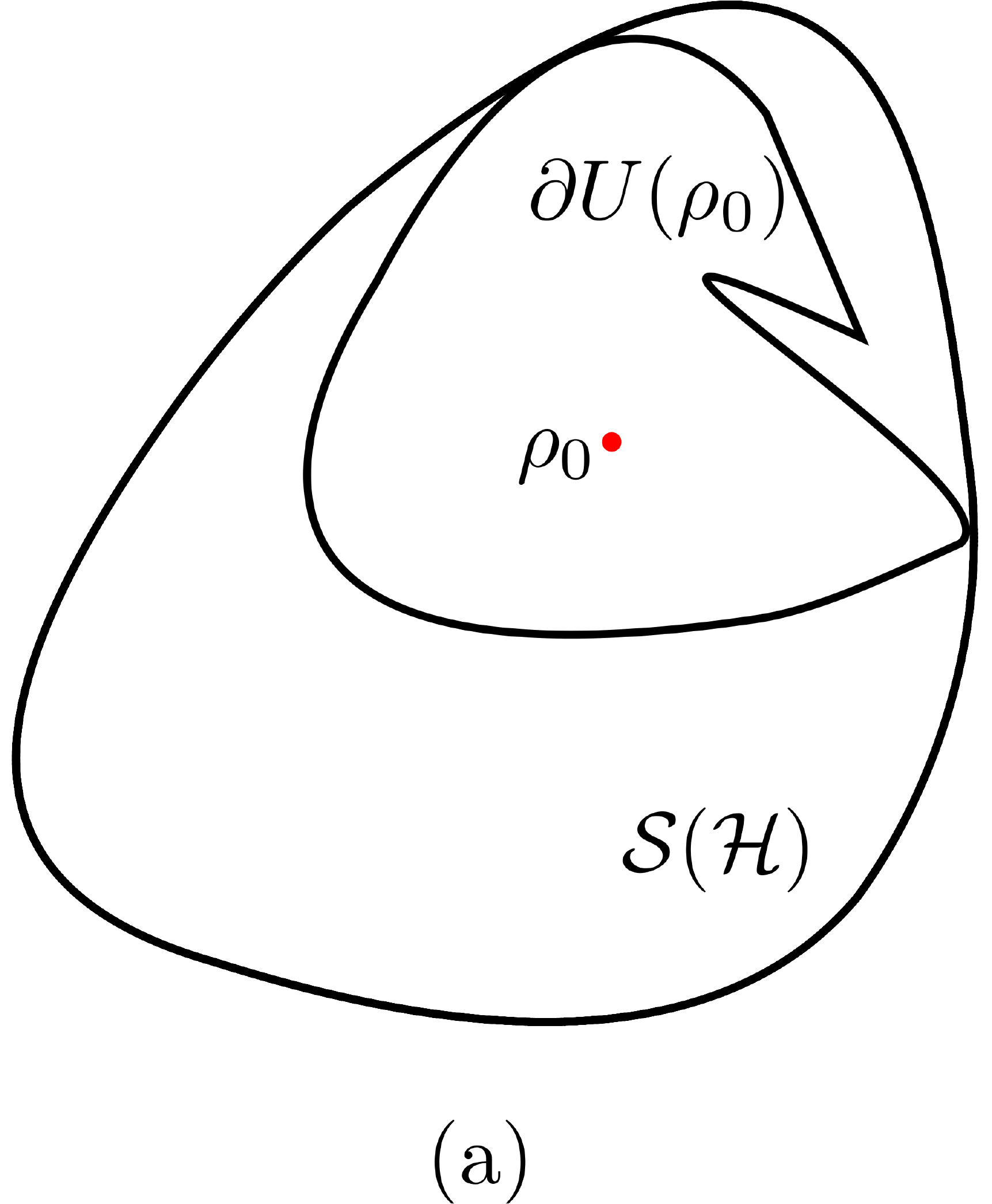}
	\hspace{0.5cm}
	\includegraphics[width=0.39\textwidth]{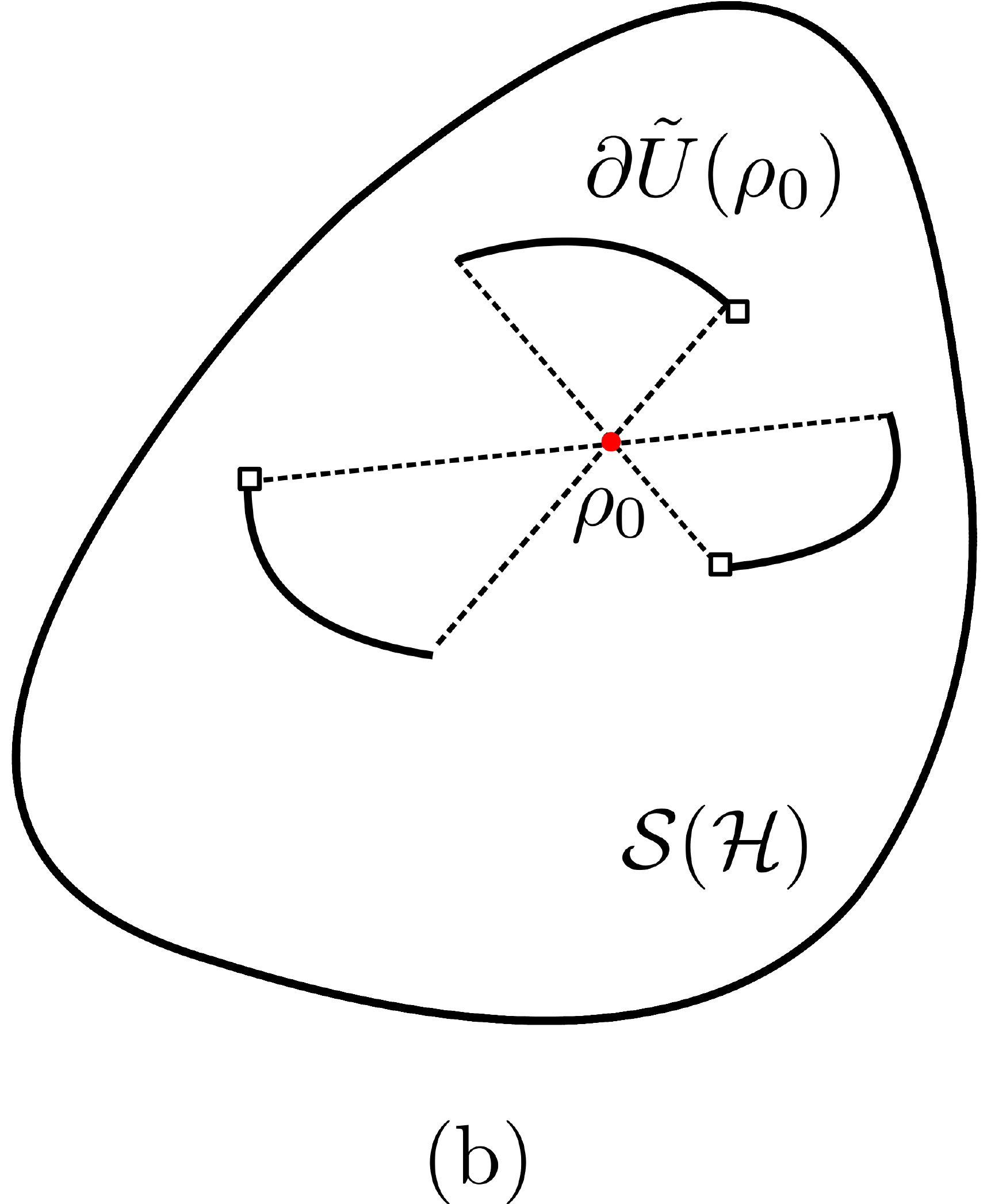}
 \end{minipage}
\caption{(Color online) Illustration of an enclosing surface $\partial U(\rho_0)$ (a) 
and of a hemispherical enclosing surface $\partial\tilde U(\rho_0)$ with 
disconnected boundary (b) for an interior point $\rho_0$ of the state space 
$\mS$.}
\label{fig:enclosing neighbourhood}
\end{figure}

The theorem bears several important mathematical and physical consequences.
First, it demonstrates that the non-Markovianity measure can be determined by
maximization over single quantum states $\rho$ taken from an
arbitrary neighborhood of a fixed state $\rho_0$ in the interior of the state 
space. Thus, Eq.~\eqref{eq:new measure} provides a {\textit{local}} 
representation of non-Markovianity, showing that quantum memory effects
can be detected locally by sampling single states from an arbitrary
enclosing surface of a fixed reference state. Note that the theorem cannot be
applied to infinite dimensional Hilbert spaces since 
$\mathring{\mathcal{S}}(\mH)$ is empty in this case.

Second, the choice of the fixed reference state $\rho_0$ is completely
arbitrary, the only condition being that it belongs to the interior of the state space.
Thus, the non-Markovianity of a dynamical process is indeed a {\textit{universal}} 
feature which appears everywhere in state space: The information about 
non-Markovian behavior is contained in any part of the state space which
supports the intuitive idea that quantum memory effects represent an intrinsic 
property of the dynamical process. This fact is particularly relevant when dealing 
with a dynamical process that has an invariant state in the interior of the
state space. It is then of great advantage to choose $\rho_0$ as this invariant state 
such that only the sampled states $\rho\in\partial U(\rho_0)$ evolve nontrivially in 
time.

Third, the theorem greatly simplifies the analytical, numerical or experimental
determination of the non-Markovianity measure. In particular, it shows that it
is not necessary to scan the whole state space in order to find an optimal pair of 
quantum states but rather sample the states of an enclosing surface
of a fixed interior point of the state space. From the proof of the theorem
we also see that it suffices if the enclosing surface contains all
directions emanating from the fixed reference state $\rho_0$ exactly once,
i.e. if Eq.~\eqref{def-U} holds for exactly one $\lambda>0$. It is even
sufficient if this equation holds for either $A$ or $-A$. Therefore, the theorem is
also valid if $\partial U(\rho_0)$ is replaced by a hemispherical enclosing surface
$\partial\tilde{U}(\rho_0)$ which we define as follows. A set 
$\partial\tilde{U}(\rho_0)\subset\mS$ is said to be a hemispherical enclosing 
surface of $\rho_0$ if and only if for any $A\in\mE_0(\mH)$ there exists exactly 
one real number $\lambda>0$ such that either 
$\rho_0+\lambda A\in\partial\tilde U(\rho_0)$ or 
$\rho_0-\lambda A\in \partial\tilde U(\rho_0)$.
A hemispherical enclosing surface thus contains all directions, given by operators $A\in\mE_0(\mH)$, only once. Moreover, it needs neither be smooth nor connected (see Fig. \ref{fig:enclosing neighbourhood}(b) for an example) which makes this characterization particularly useful for noisy experiments.

\begin{figure}[tbh]
 \centering
 \includegraphics[width=0.39\textwidth]{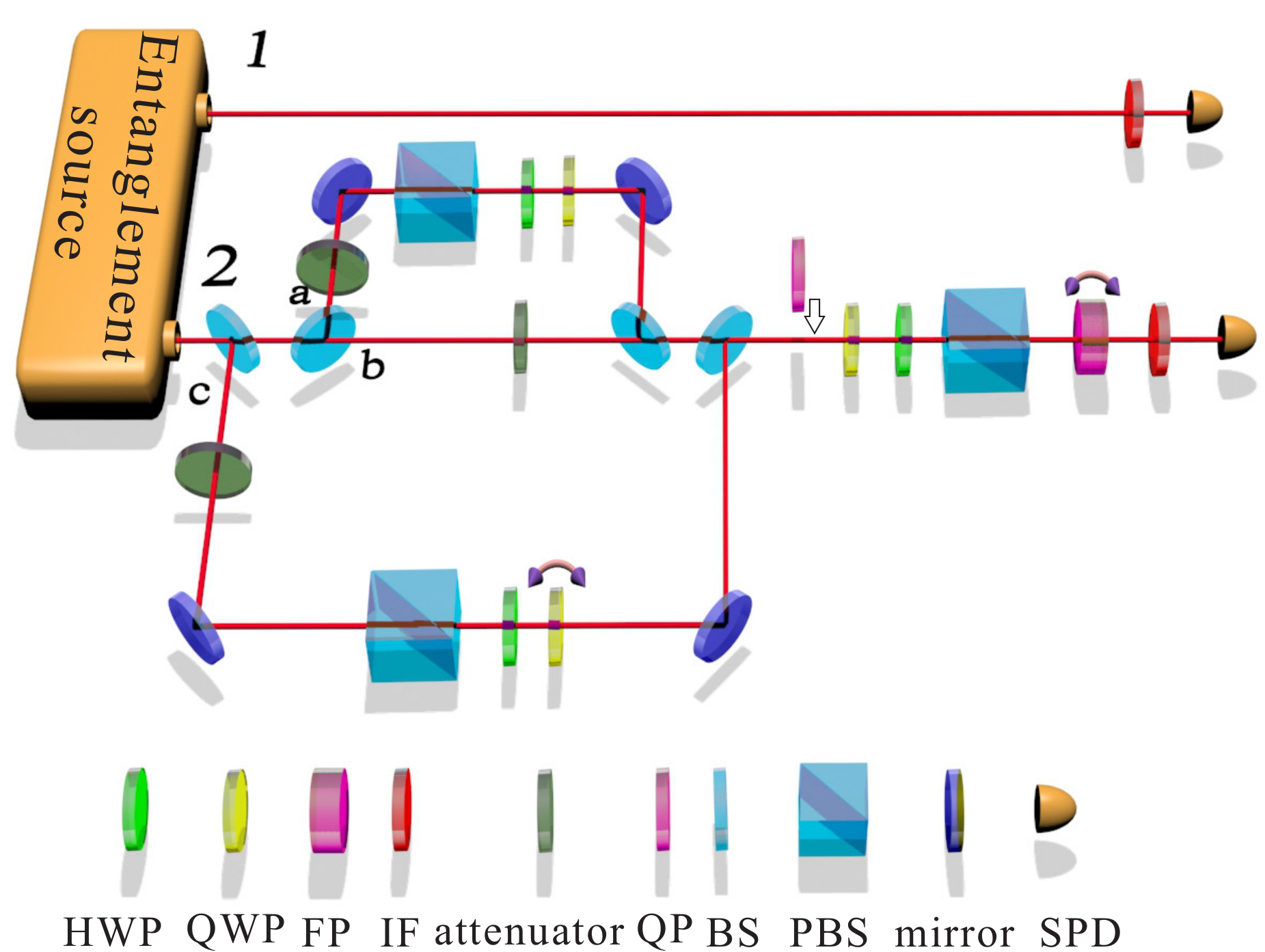}
 \caption{(Color online) Experimental setup. Key to the components: HWP -- half-wave plate, QWP -- quarter-wave plate, FP -- Fabry-P\'erot cavity, IF -- interference filter, QP -- quartz plate, (P)BS -- (polarizing) beamsplitter, SPD -- single photon detector.}\label{fig:setup}
\end{figure}

We have applied the above theorem to determine the degree of non-Markovianity 
in a photonic process. The open quantum system is provided by the polarization degree of freedom of photons coupled to the frequency degree representing the environment. The experimental setup is depicted in Fig. \ref{fig:setup}. With the help of a frequency doubler a mode-locked Ti:sapphire laser (central wavelength $780$\,nm) is used to pump two 
$1$\,mm thick BBO crystals to generate the maximally entangled two-qubit state 
$(|H,V\rangle-|V,H\rangle)/\sqrt{2}$ with $|H\rangle$ and $|V\rangle$ denoting the horizontal and vertical polarization states, respectively \cite{Niu08}. A fused silica plate ($0.1$\,mm thick and coated with a partial reflecting coating, with approximately $80\,$\% reflectivity at $780$\,nm) serves as a Fabry-P\'erot cavity (FP) which in addition can be tilted to generate different dynamical behavior 
\cite{exp2}. The cavity and a consecutively placed interference filter (IF) (FWHM about $3$\,nm) single out two peaks near $780$\,nm of width $\sigma=7.7\times 10^{11}$\,Hz each which are separated by $\Delta\omega=7.2\times 10^{12}$\,Hz. The relative amplitude $A_\alpha$ of the two peaks depends strongly on the tilt angle $\alpha$ whereas the other quantities are almost constant.
A polarizing beamsplitter (PBS) together with a half-wave plate (HWP) and a quarter-wave plate (QWP) are used as a photon state analyzer \cite{James01}.

Photon $1$ is directly detected in a single photon detector at the end of arm $1$ as a trigger for photon $2$. The optical setup in part $a$, $b$ and $c$ (see Fig. \ref{fig:setup}) is used to prepare arbitrary quantum states of photon $2$ needed for the sampling process \cite{Kwiat99}. This set-up conveniently allows to prepare any single pure photon polarization state (in arm $2c$) and reference states ($2a$ along with $2b$) together with arbitrary enclosing surfaces which can be controlled by changing the relative amplitudes of the attenuators built in in each arm. The path difference between each arm is about $25$\,mm to ensure that the mixture of the three parts is classical.


After the preparation photon $2$ passes through birefringent quartz plates of 
variable thickness which couple the polarization and frequency degree of freedom 
and lead to the decoherence of superpositions of polarization states. The 
birefringence is given by $\Delta n=8.9\times10^{-3}$ at $780$\,nm. 
The thickness of the 
quartz plates simulating different evolution times ranges from $75\lambda$ to 
$318\lambda$ in units of the central wavelength of the FP cavity.

Employing the Bloch vector representation, the set of polarization states can be 
conveniently parametrized by means of spherical coordinates 
$\mathbf{r}=(r,\theta,\phi)$. 
We apply the local representation to two reference states to determine experimentally the degree of non-Markovianity for three dynamics characterized by the relative amplitudes $A_\alpha=0.64$, $0.22$ and $0.01$, ranging from non-Markovian to Markovian evolutions, and compare the results with the outcome for pairs of orthogonal initial states. The reference states $\rho_0^1$ and $\rho_0^2$ used in the experiment are given by
\begin{align}
 \mathbf{r}_0^{1}=\bigl(0.20,\tfrac{1}{2}\pi,\tfrac{13}{50}\pi\bigr),
 \qquad
 \mathbf{r}_0^2=\bigl(0.88,\tfrac{8}{50}\pi,\tfrac{13}{50}\pi\bigr).
\end{align}
Reference state $\rho_0^1$ is thus located inside the equatorial plane, whereas the second reference state lies in the northern hemisphere close to the boundary. The enclosing surfaces are determined by the convex combination $ 0.3\cdot\rho_0^{a,b}+0.7\cdot\rho$ of the reference states and any pure state prepared in arm $2c$. These sets thus contain only mixed states. We measured a total of $5000$ states on the surface for each reference state which are characterized by the azimuthal and polar angles of the pure states. The associated angles $\theta$ and $\phi$ are located on a lattice with equal spacing of $2\pi/100$. 
 
\begin{figure}[tbh]%
    \centering
    \subfloat[]{{\includegraphics[width=2.8cm]{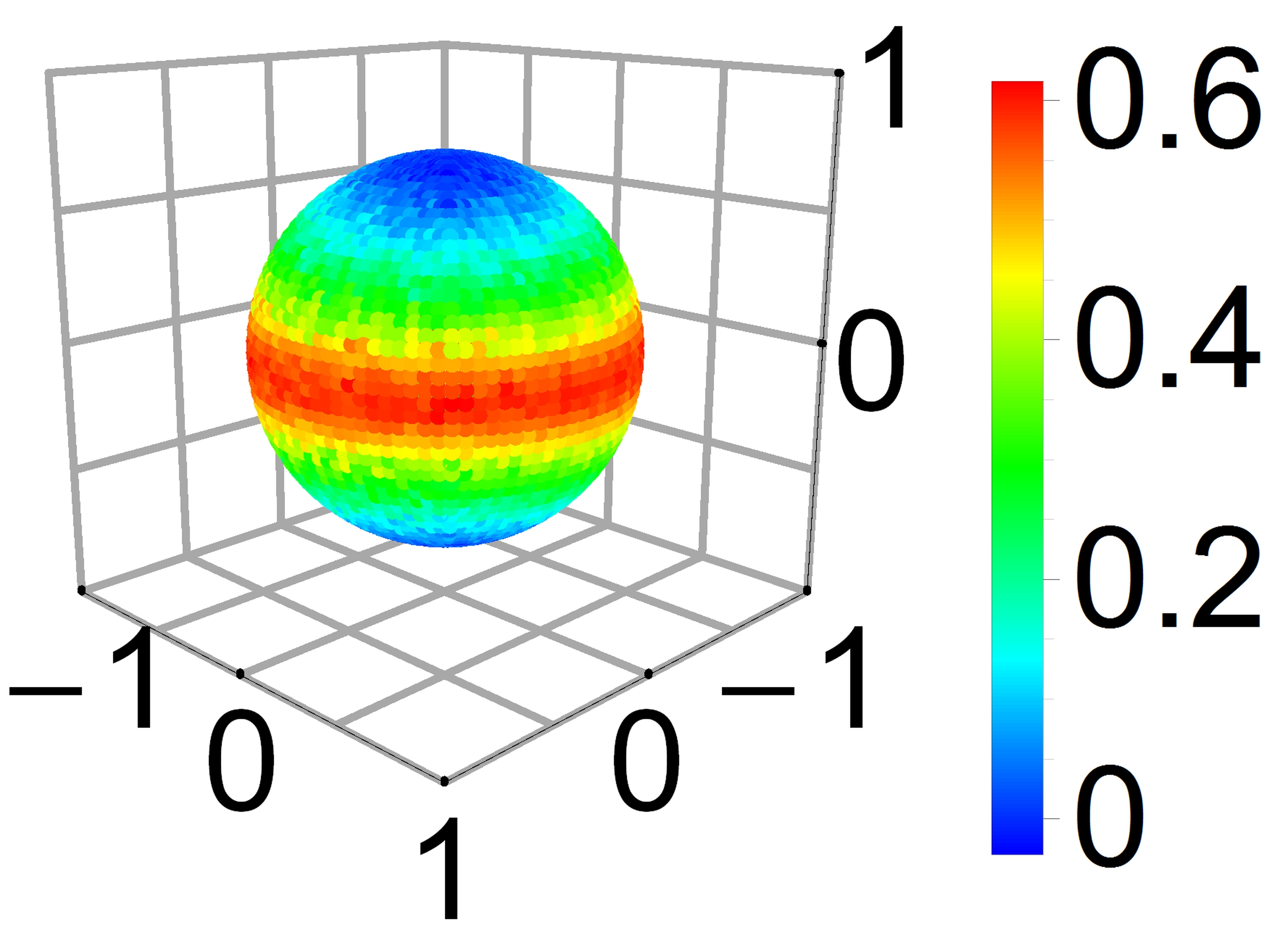} }}
    \subfloat[]{{\includegraphics[width=2.8cm]{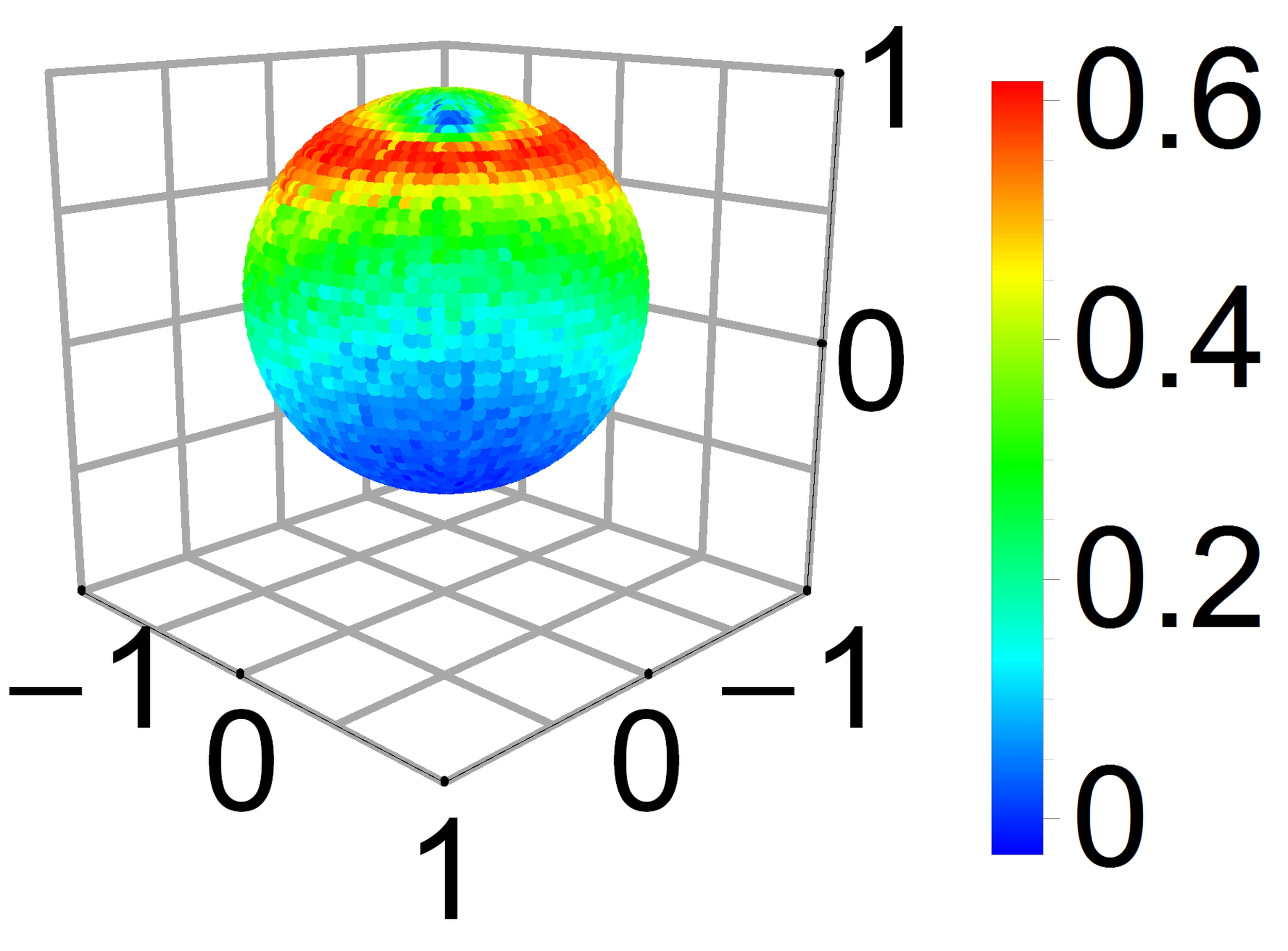} }}
    \subfloat[]{{\includegraphics[width=2.8cm]{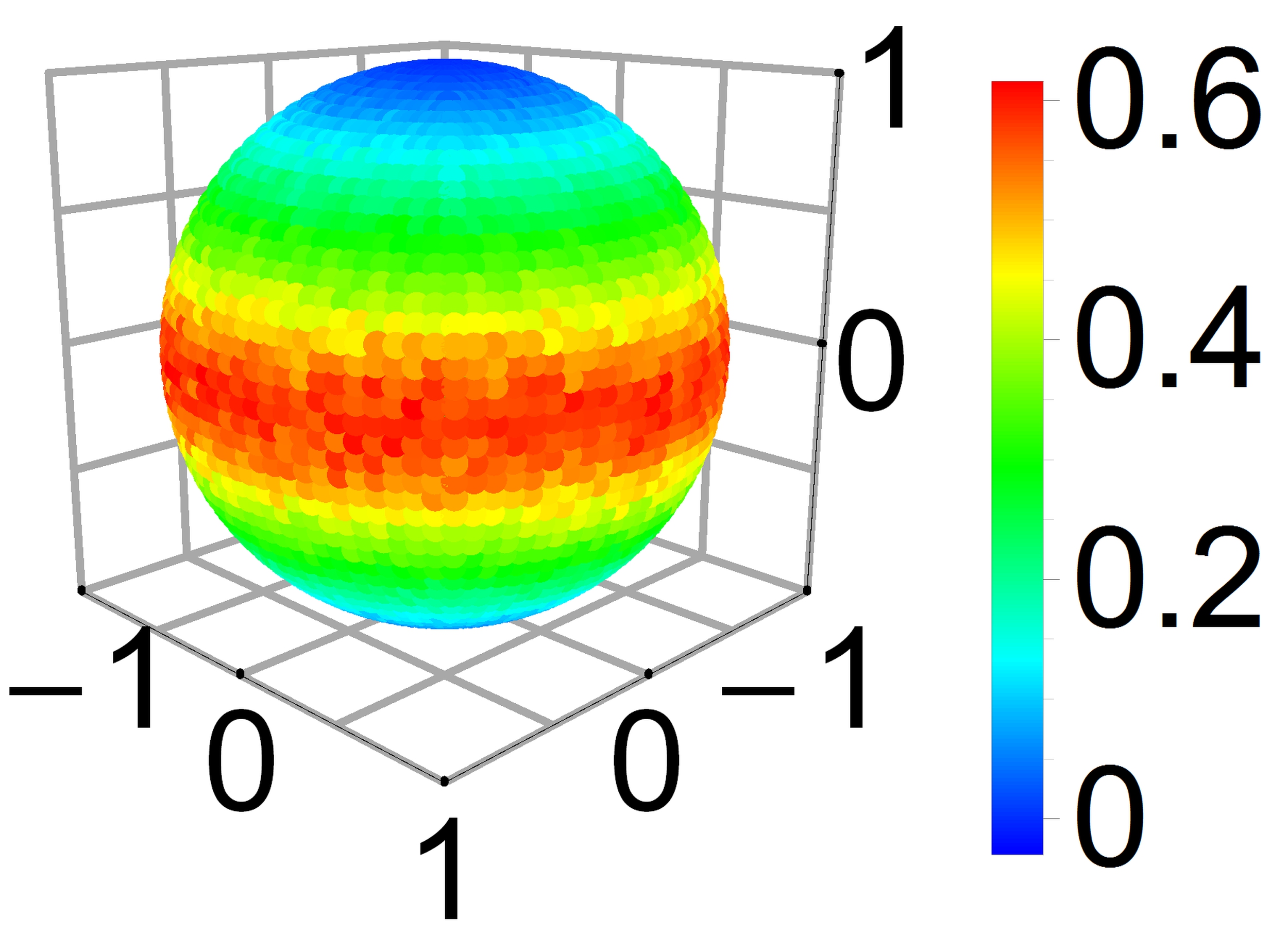} }}
    \vspace{-0.2cm}
    \subfloat[]{{\includegraphics[width=2.8cm]{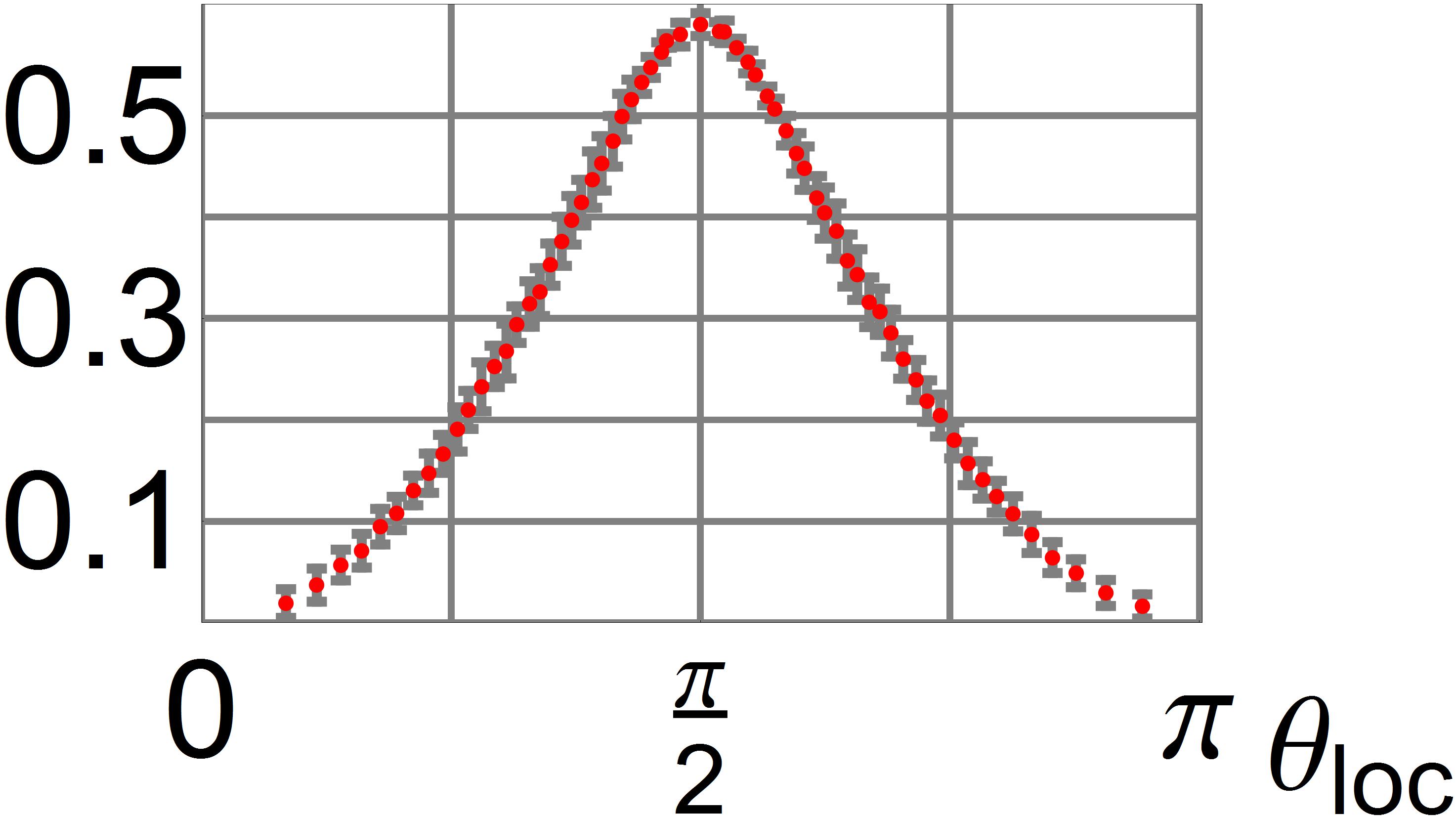} }}
    \subfloat[]{{\includegraphics[width=2.8cm]{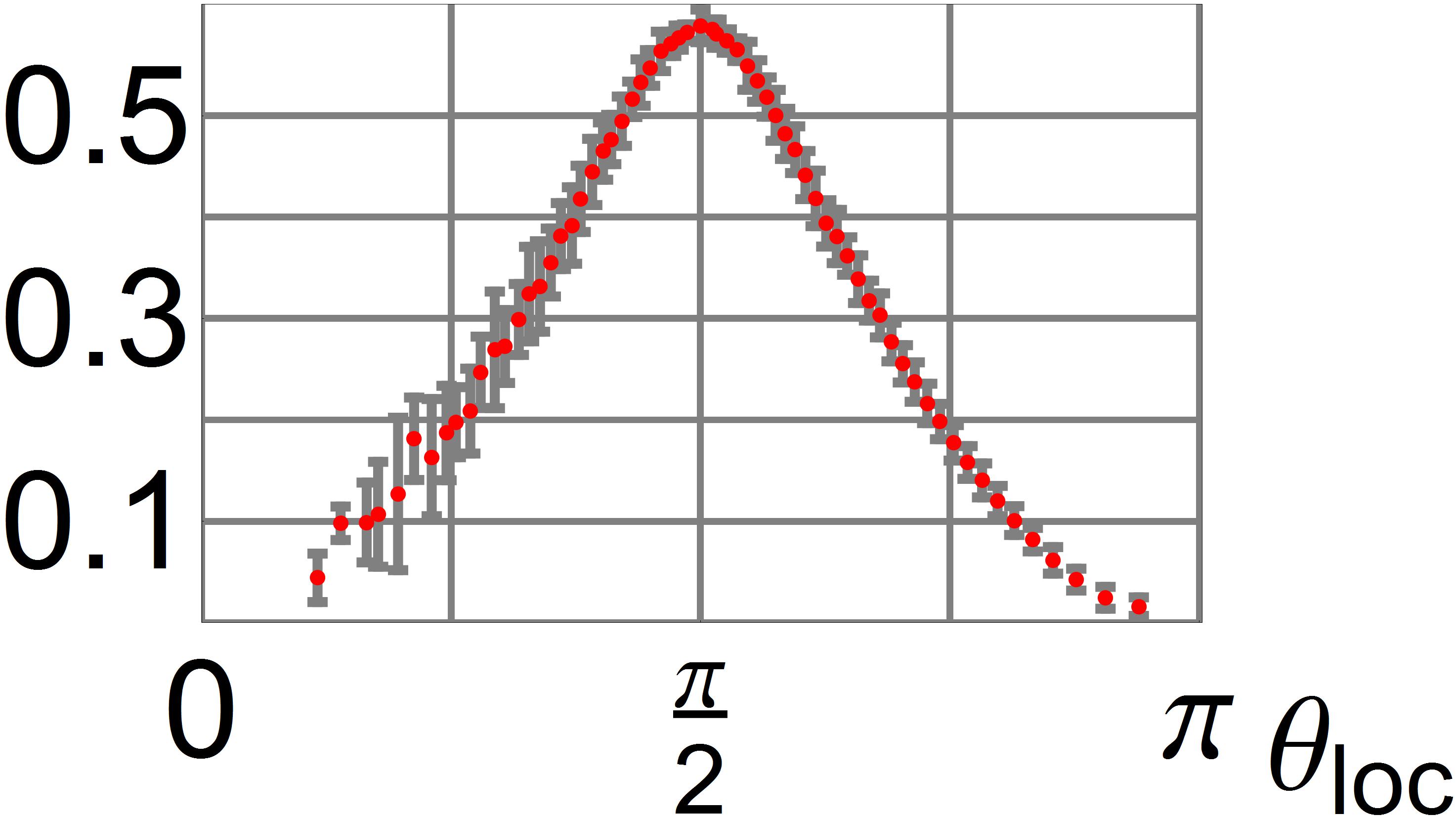} }}
    \subfloat[]{{\includegraphics[width=2.7cm]{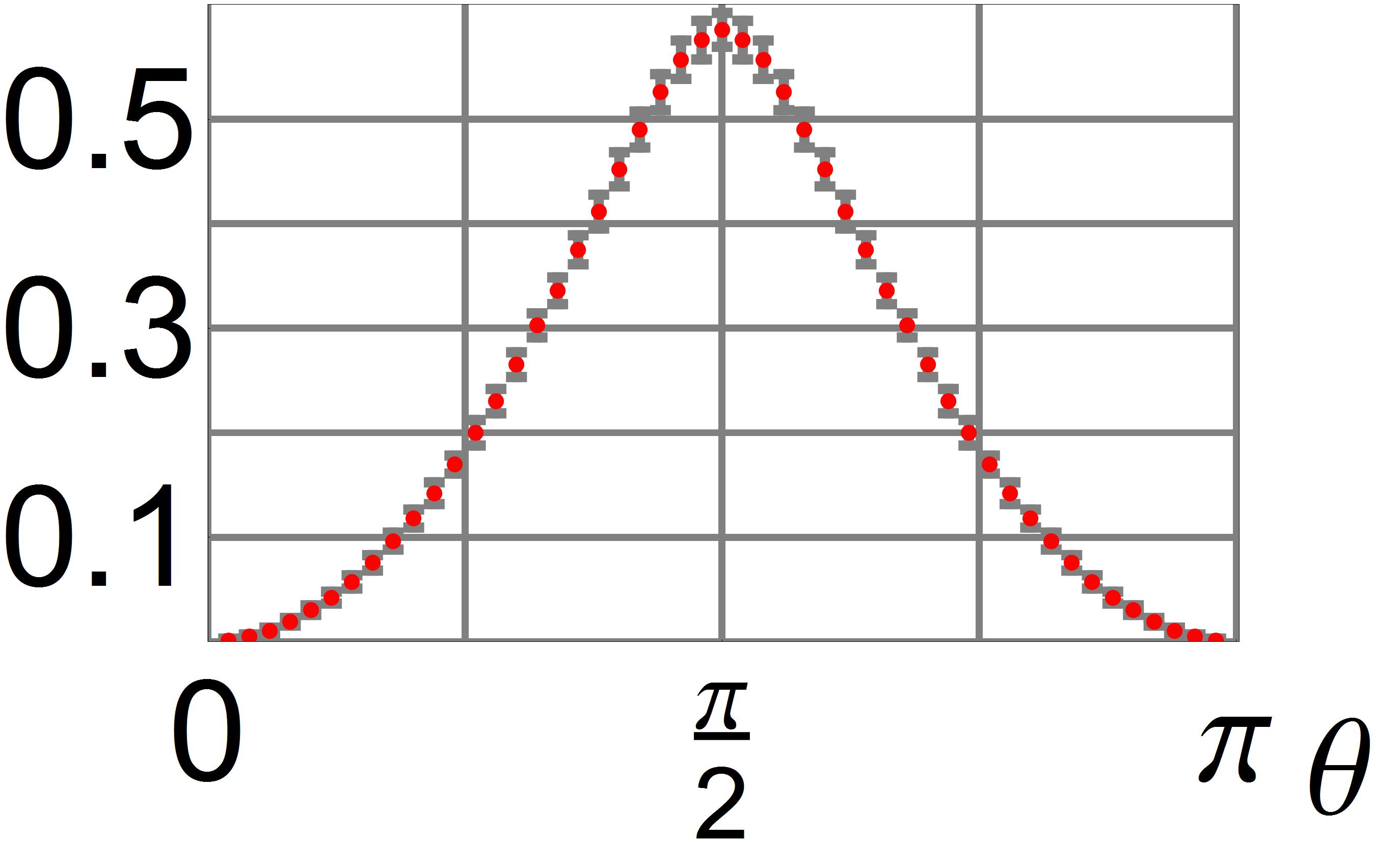} }}
    \caption{(Color online) Experimental results for the increase of the trace distance between $175\lambda$ and $318\lambda$ for $A_\alpha=0.64$ for states on the enclosing surface of reference state $\rho_0^1$ (a), $\rho_0^2$ (b) and pairs of orthogonal states (c). The corresponding $\phi_\mathrm{loc}$-averaged increase with respect to local spherical coordinates is shown in (d), (e) and (f). Error bars show
    the standard deviations.}
    \label{fig:N=0.59}%
\end{figure}

\begin{figure}[tbh]
    \centering
    \subfloat[]{{\includegraphics[width=2.8cm]{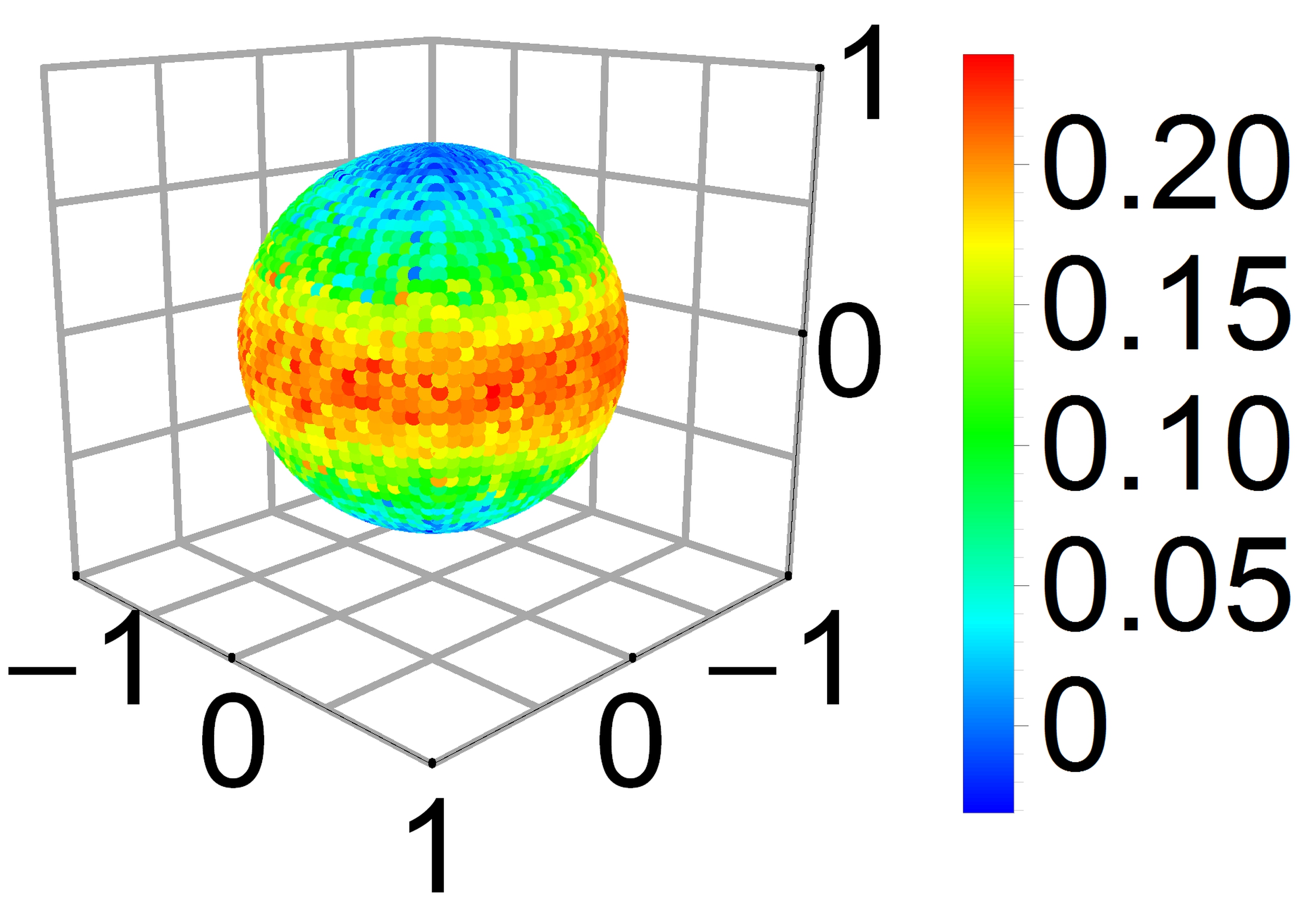} }}
    \subfloat[]{{\includegraphics[width=2.8cm]{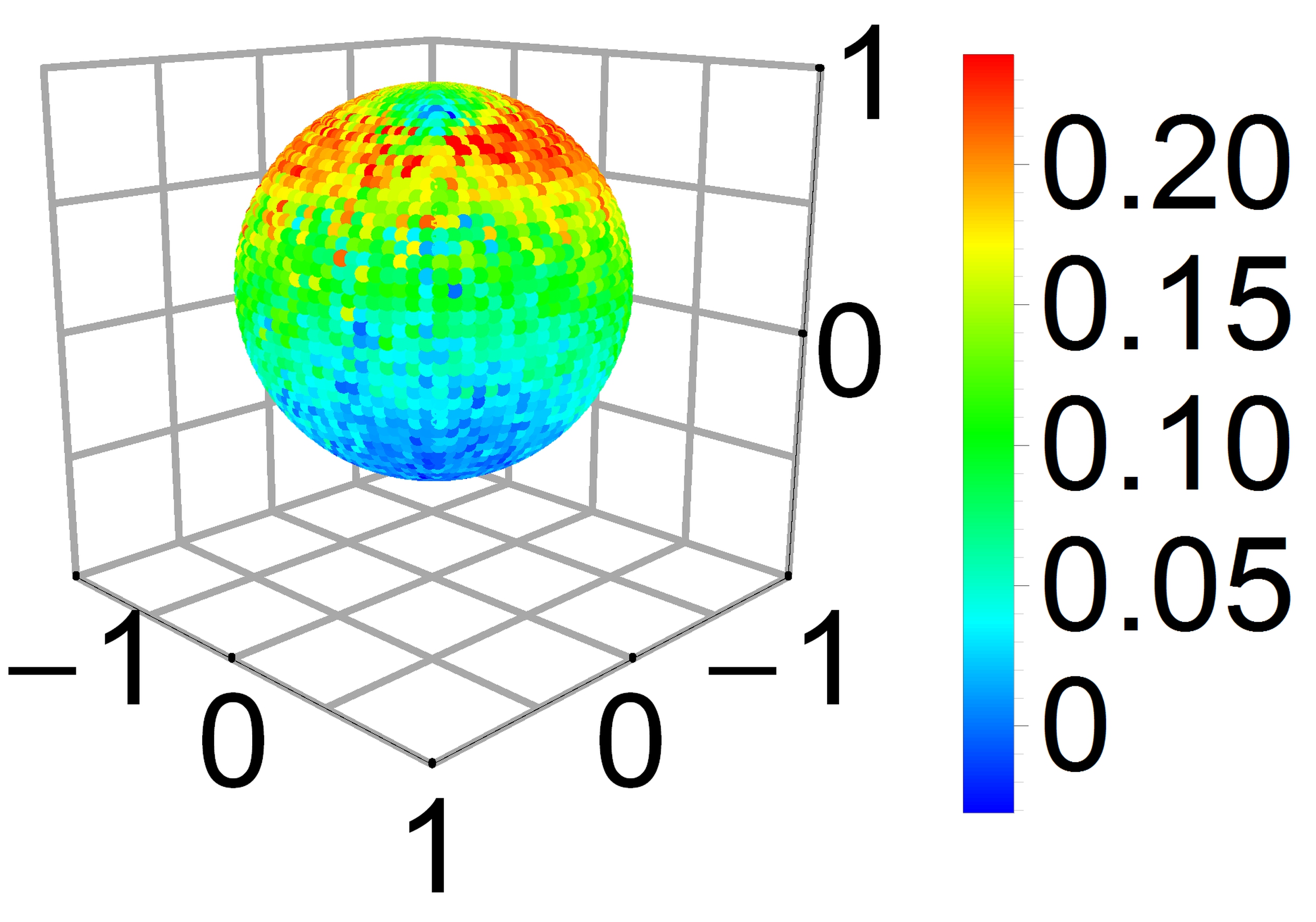} }}
    \subfloat[]{{\includegraphics[width=2.8cm]{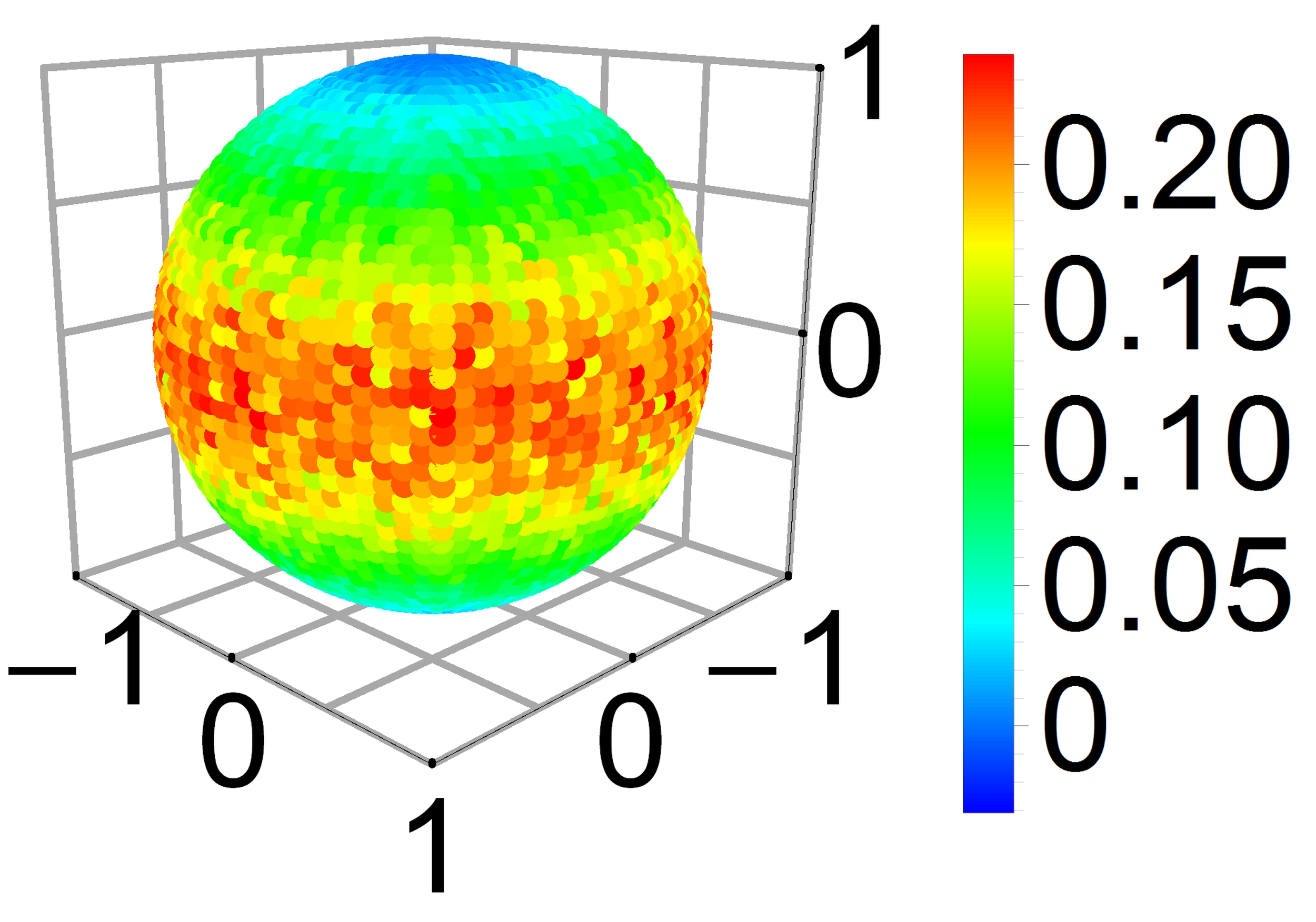} }}
    \vspace{-0.2cm}
    \subfloat[]{{\includegraphics[width=2.8cm]{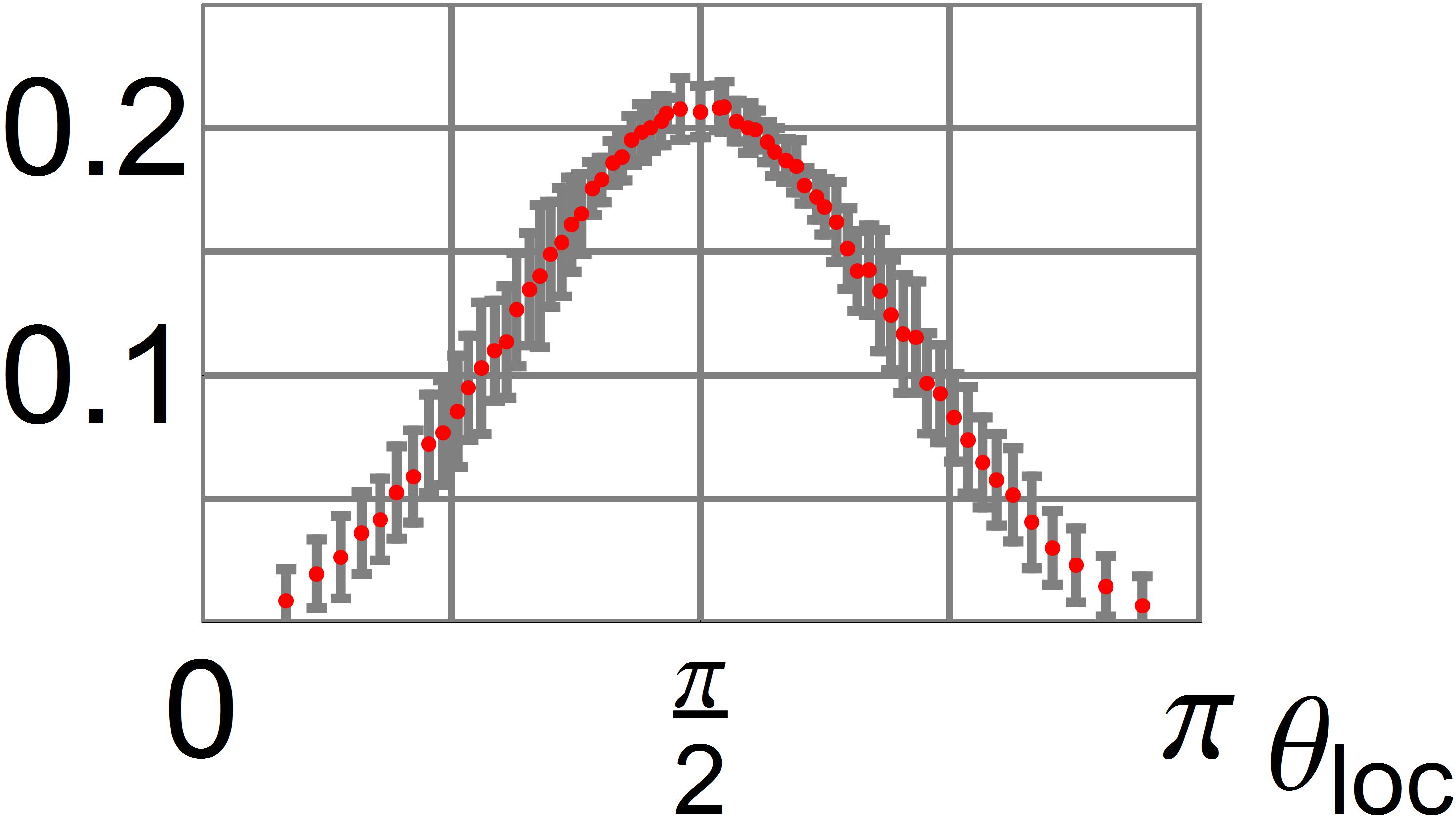} }}
    \subfloat[]{{\includegraphics[width=2.8cm]{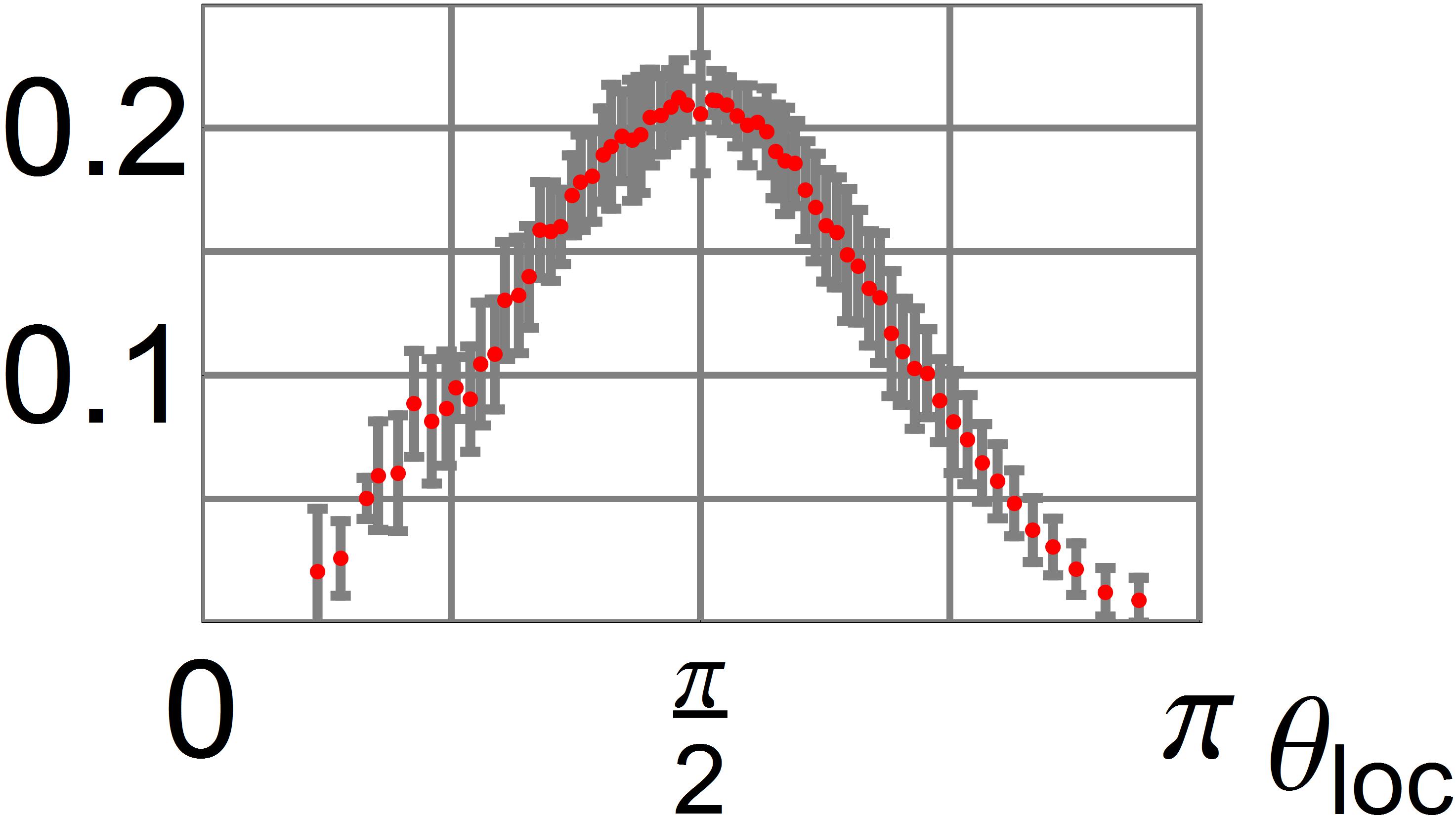} }}
    \subfloat[]{{\includegraphics[width=2.7cm]{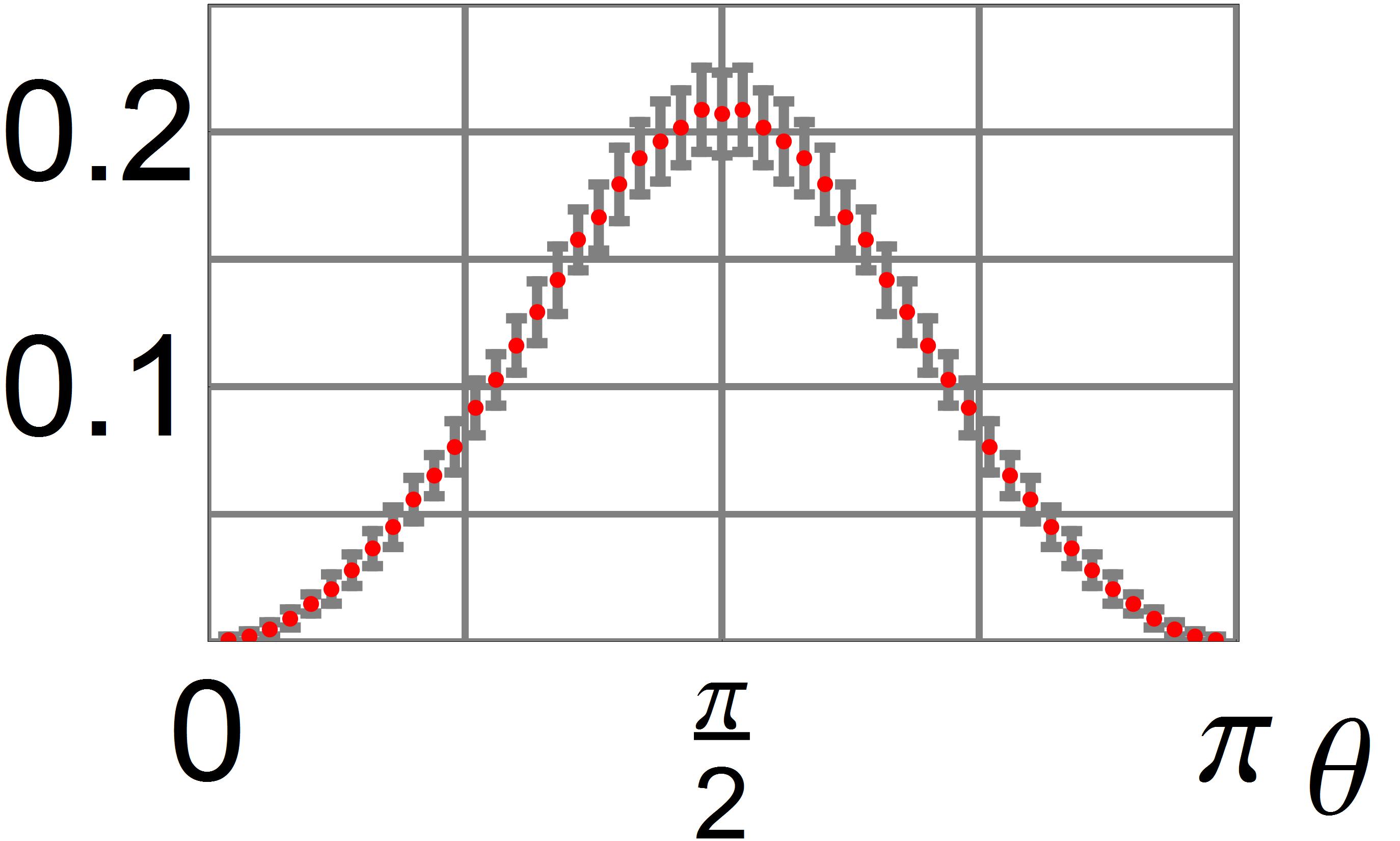} }}
\caption{(Color online) The same as Fig.~\ref{fig:N=0.59} for $A_\alpha=0.22$.}
    \label{fig:N=0.21}
\end{figure}

\begin{figure}[tbh]
    \centering
     \subfloat[]{{\includegraphics[width=2.8cm]{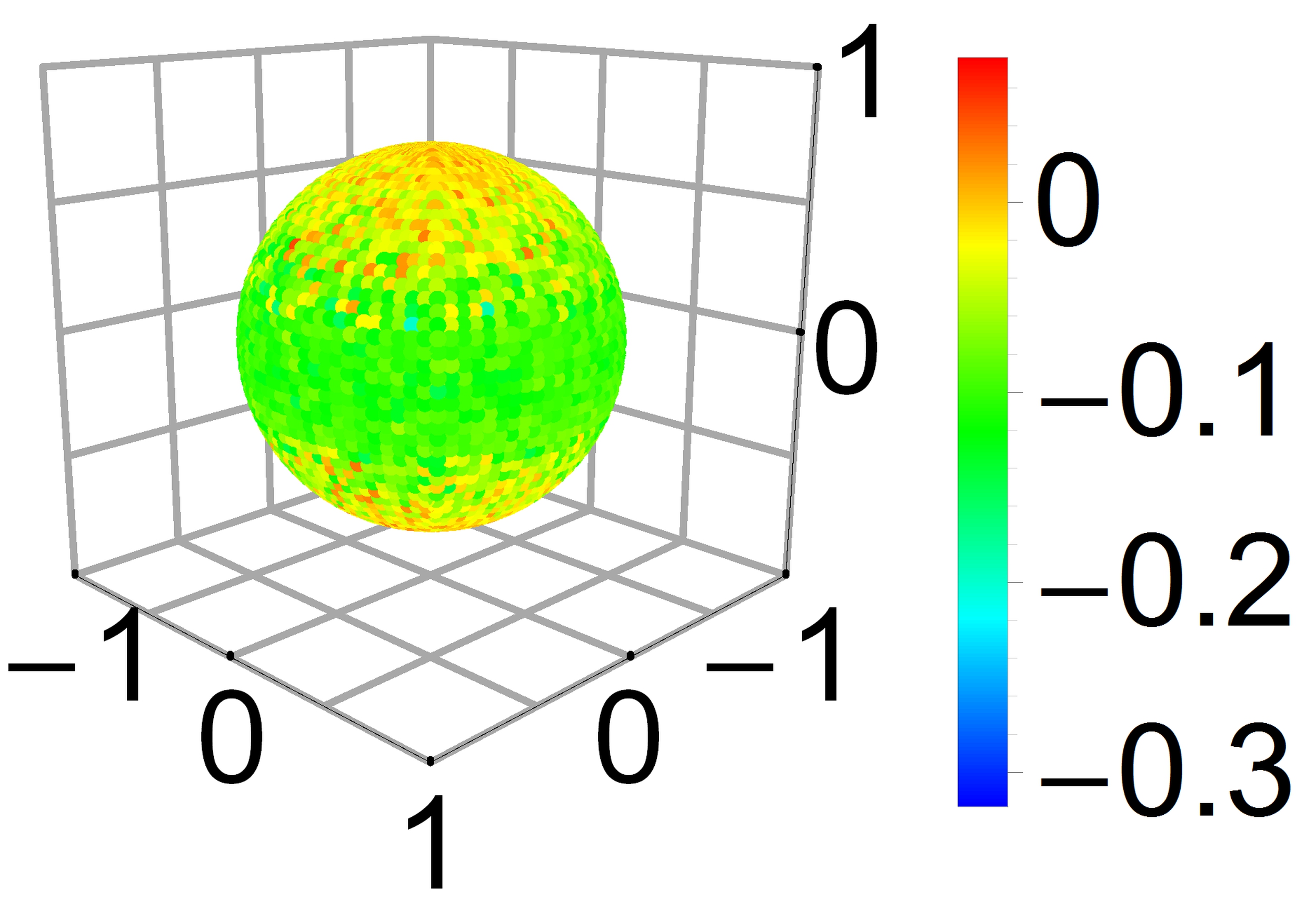} }}
     \subfloat[]{{\includegraphics[width=2.8cm]{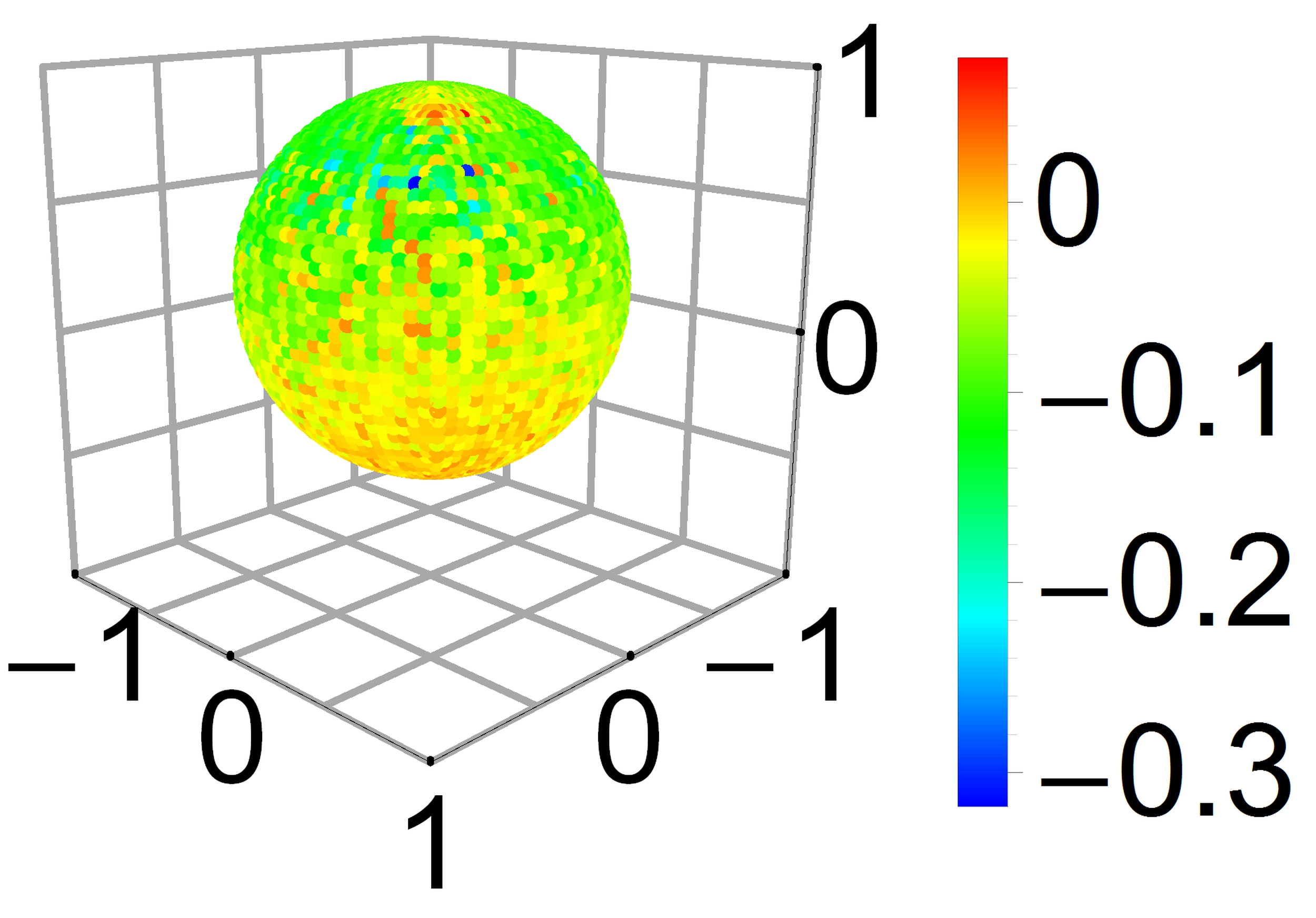} }}
     \subfloat[]{{\includegraphics[width=2.8cm]{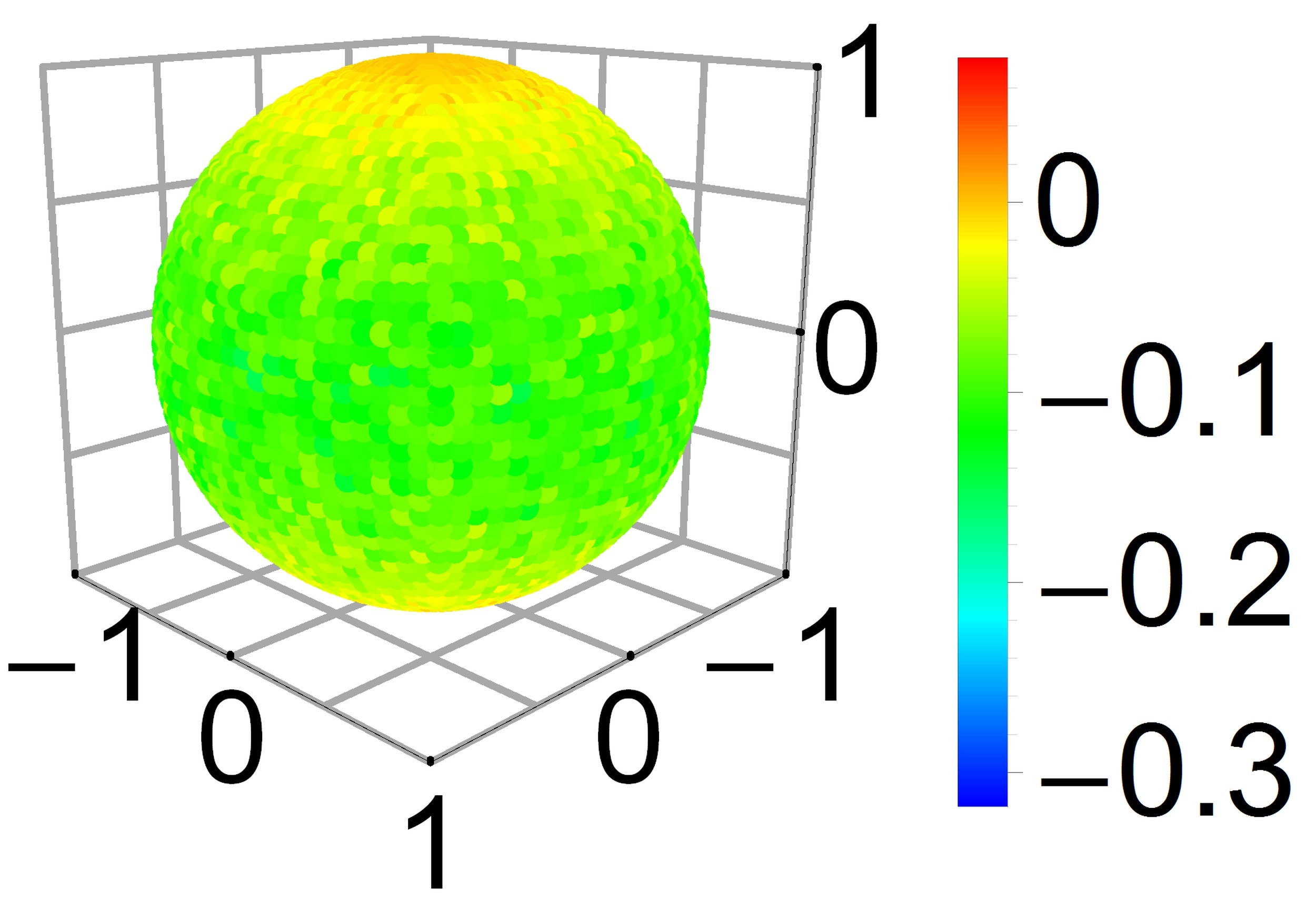} }}
     \vspace{-0.2cm}
     \subfloat[]{{\includegraphics[width=2.8cm]{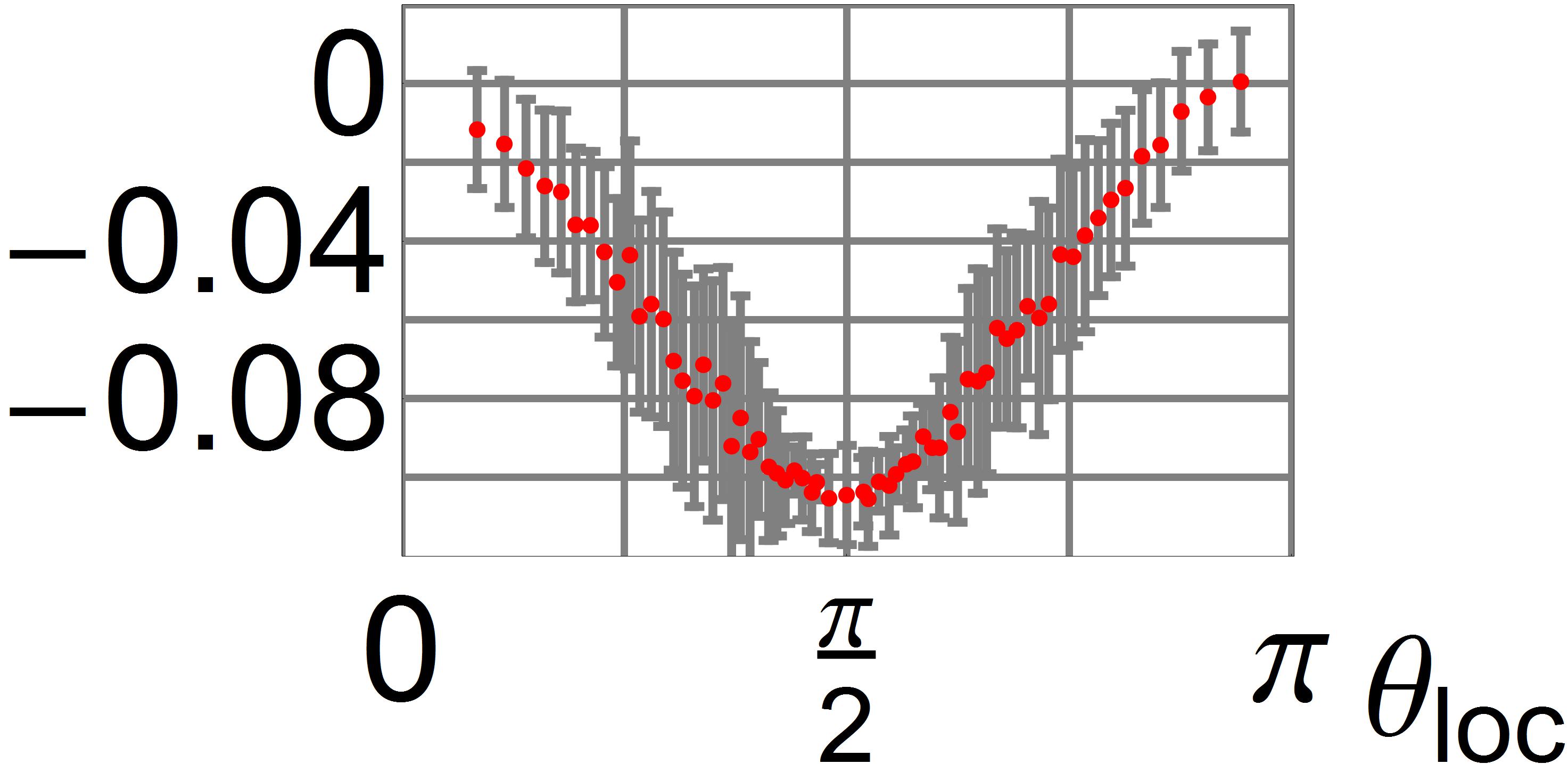} }}
     \subfloat[]{{\includegraphics[width=2.8cm]{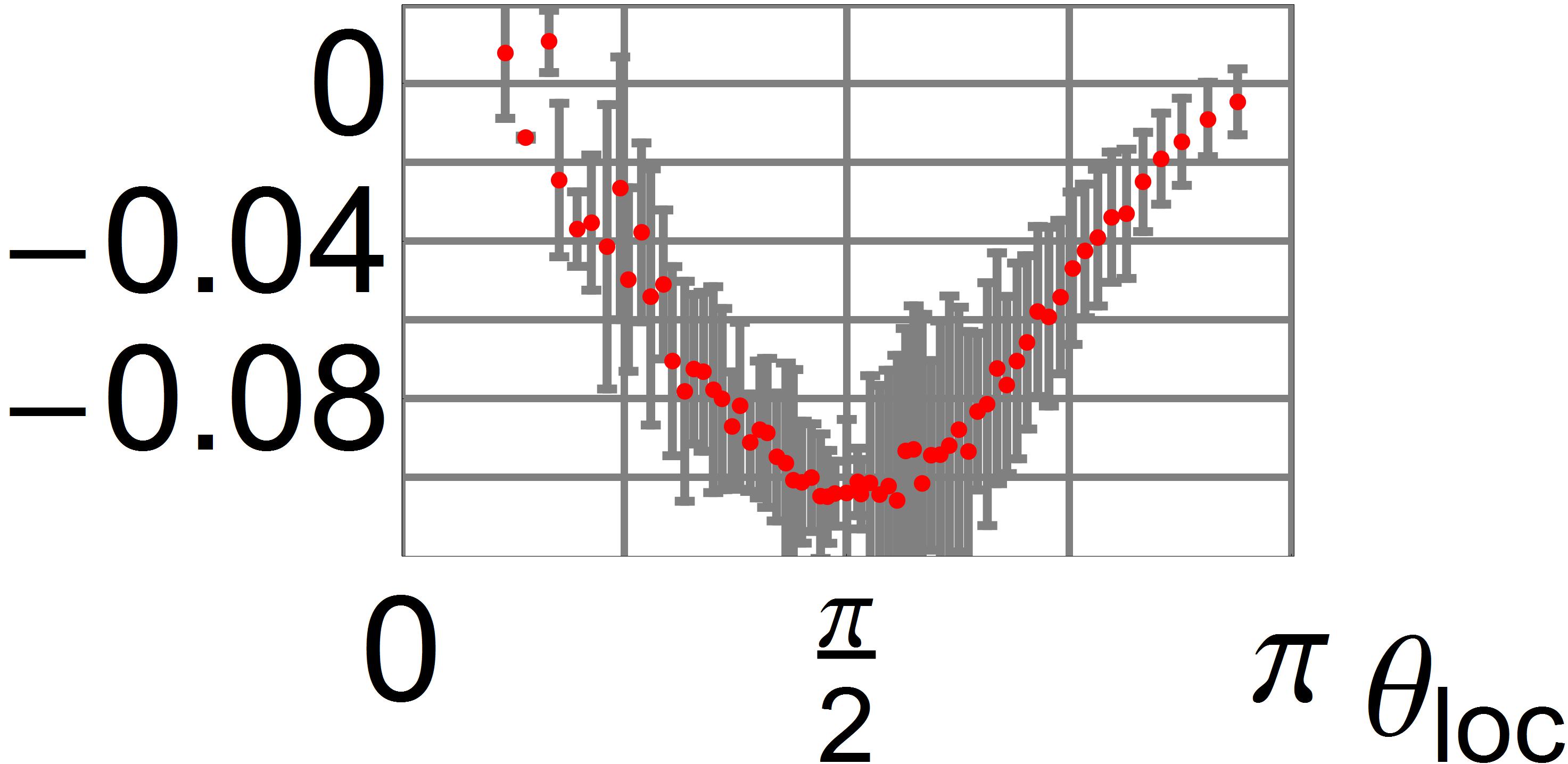} }}
     \subfloat[]{{\includegraphics[width=2.7cm]{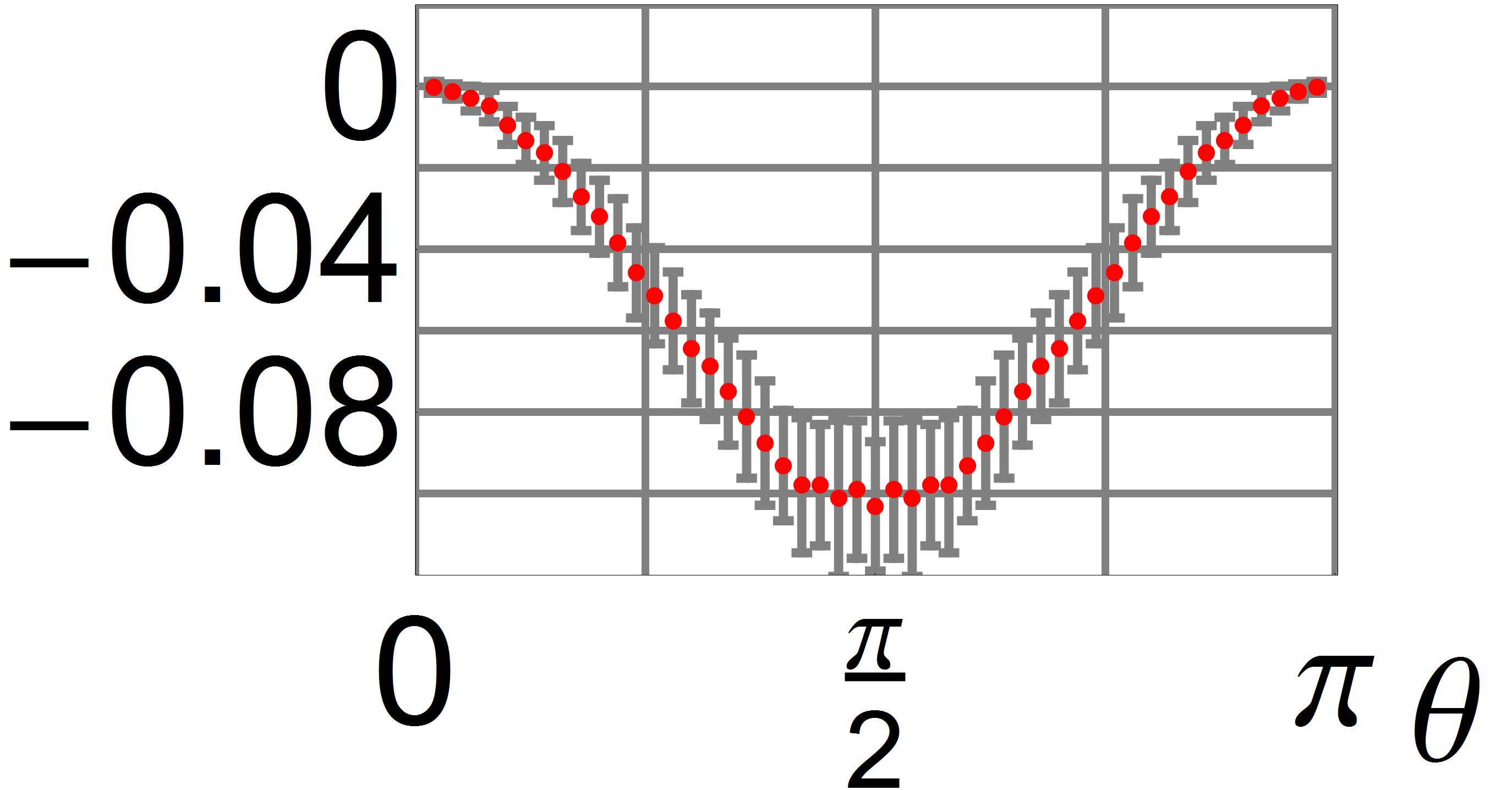} }}
\caption{(Color online) The same as Fig.~\ref{fig:N=0.59} for $A_\alpha=0.01$.}
     \label{fig:N=0}
 \end{figure}

The outcomes of the measurements are presented in Figs.~\ref{fig:N=0.59}, 
\ref{fig:N=0.21} and \ref{fig:N=0}. The increase of the trace distance between 
$175\lambda$ and $318\lambda$ for any state on the enclosing surface for the 
two reference states is shown in Figs.~\ref{fig:N=0.59}(a)-\ref{fig:N=0}(a) and 
\ref{fig:N=0.59}(b)-\ref{fig:N=0}(b) using color coding. Note, that the colored 
surfaces in these figures are non-spherical and not centered at the origin.
By contrast, the ordinary Bloch spheres depicted in 
Figs.~\ref{fig:N=0.59}(c)-\ref{fig:N=0}(c) show the measurement outcomes
for pairs of orthogonal initial states.

Defining spherical coordinates 
$\mathbf{r}_\mathrm{loc}=(r_\mathrm{loc},\theta_\mathrm{loc},\phi_\mathrm{loc})$ 
with respect to local coordinate systems centered at the position of the two
reference states $\rho_0^1$ and $\rho_0^2$, 
one recovers the polar symmetry present for pairs of orthogonal states, see
Figs.~\ref{fig:N=0.59}(a)-(c).
One may therefore average the outcomes over the polar angle $\phi_\mathrm{loc}$ along lines of latitude. To this end, we introduced an appropriate binning on the $z$-axis and determined the average increase which we then assigned to the azimuthal angle $\theta_\mathrm{loc}$ associated to the mean $z$-value in the bin. In addition, we allocated the standard deviation to each of the averaged outcomes. The resulting data are depicted in Figs.~\ref{fig:N=0.59}(d)-\ref{fig:N=0}(d) and \ref{fig:N=0.59}(e)-\ref{fig:N=0}(e) and show the same characteristics as the $\phi$-averaged increase of pairs of orthogonal states displayed in Figs.~\ref{fig:N=0.59}(f)-\ref{fig:N=0}(f). Note that the directional dependence of the trace distance originating from its property of depending only on the difference of two states can be nicely seen for example in Figs.~\ref{fig:N=0.59}(d)-(f).

The maximal increase of the trace distance for the two reference states obtained 
from the $\phi_\mathrm{loc}$-averaged data as well as for pairs of orthogonal 
states are given in Tab.~\ref{tab:results}. The experimentally determined values 
are in very good agreement with the predictions of the theoretical model 
\cite{exp2}, demonstrating the experimental feasibility and the accuracy of the 
method based on the local representation \eqref{eq:new measure}.

Summarizing, we have derived a representation of the measure for quantum 
non-Markovianity which fully reveals the locality and the universality of this
measure. These properties are clearly reflected in the results of our photonic 
experiment. The experiment illustrates that the measure can be obtained 
efficiently in an arbitrary neighborhood of any fixed state in the interior of the 
state space, that its determination only requires a maximization over a single
input state, and that optimal quantum states featuring maximal backflow of
information can always be represented by mixed states.

\begin{table}[tbh]
\begin{center}
\begin{tabular}{|c||c|c|c|c|}
	\hline
$A_\alpha$&$\mathcal{N}_\mathrm{theo}$	   &$\mathcal{N}_\mathrm{(a)}$&$\mathcal{N}_\mathrm{(b)}$&$\mathcal{N}_\mathrm{(c)}$ \\ \hline\hline
$0.64$		& $0.59$	&$0.59\pm0.01$		&$0.59\pm0.02$		&$0.59\pm0.02$		\\
$0.22$		& $0.21$	&$0.21\pm0.01$		&$0.21\pm0.02$		&$0.21\pm0.02$		\\
$0.01$		& $0$		&$0.001\pm0.013$	&$-0.005\pm0.008$	&$-0.0002\pm0.0015$	\\
\hline
\end{tabular}
\caption{The quantum non-Markovianity measure for the three dynamics obtained 
from the experimental data in comparison to the theoretical value.}
\label{tab:results}
\end{center}
\end{table}

\acknowledgments
This work was supported by the National Basic Research Program of China 
(2011CB921200), the CAS, the National Natural Science Foundation of China 
(11274289, 11325419,11374288, 11104261, 61327901), the National Science 
Fund for Distinguished Young Scholars (61225025), the Fundamental Research 
Funds for the Central Universities (WK2470000011), the Academy of Finland 
(Project 259827), the Jenny and Antti Wihuri Foundation, the Magnus Ehrnrooth 
Foundation, and the German Academic Exchange Service (DAAD). S.~W. thanks 
the German National Academic Foundation for support.

\end{document}